\documentclass[twocolumn,final,prb,showpacs,superscriptaddress,floatfix,amsmath,amssymb,amsfonts]{revtex4-1}

\usepackage{graphicx}
\usepackage{multirow}
\usepackage{array}
\usepackage{verbatim}

\usepackage{times}

\usepackage{amsmath}
\usepackage{amssymb}
\usepackage{amsfonts}
\usepackage{latexsym}

\usepackage{booktabs}

\usepackage{color}

\usepackage[colorlinks,bookmarks=false,citecolor=blue,linkcolor=red,urlcolor=blue]{hyperref}
\usepackage{verbatim}
\def\be{\begin{equation}}
\def\ee{\end{equation}}
\def\bea{\begin{eqnarray}}
\def\eea{\end{eqnarray}}

\begin{document}
\title{Local magnetism and structural properties of Heusler Ni$_2$MnGa alloys}

\author{M. Belesi}\email{m.e.belesi@ifw-dresden.de}
\affiliation{Leibniz Institute for Solid State and Materials Research, Helmholtzstrasse 20, 01069 Dresden, Germany}
\author{{L. Giebeler }}
\affiliation{Leibniz Institute for Solid State and Materials Research, Helmholtzstrasse 20, 01069 Dresden, Germany}
\author{C. G. F. Blum}
\affiliation{Leibniz Institute for Solid State and Materials Research, Helmholtzstrasse 20, 01069 Dresden, Germany}
\author{B. B\"{u}chner}
\affiliation{Leibniz Institute for Solid State and Materials Research, Helmholtzstrasse 20, 01069 Dresden, Germany}
\affiliation{Institute for Solid State Physics, TU Dresden, D-01069, Germany}
\author{S. Wurmehl}\email{s.wurmehl@ifw-dresden.de}
\affiliation{Leibniz Institute for Solid State and Materials Research, Helmholtzstrasse 20, 01069 Dresden, Germany}
\affiliation{Institute for Solid State Physics, TU Dresden, D-01069, Germany}

\date{\today}

\begin{abstract}
We present a detailed experimental study of bulk and powder samples of the Heusler shape memory alloy Ni$_2$MnGa, including zero-field static and dynamic $^{55}$Mn NMR experiments, X-ray powder diffraction and magnetization experiments. The NMR spectra give direct access to the sequence of structural phase transitions in this compound, from the high-T austenitic phase down to the low-T martensitic phase. In addition, a detailed investigation of the so-called {\it rf}-enhancement factor provides local information for the magnetic stiffness and restoring fields for each separate coordination, structural, crystallographic environment, thus differentiating signals coming from austenitic and martensitic components. The temperature evolution of the NMR spectra and the {\it rf}-enhancement factors shows strong dependence on sample preparation. In particular, we find that sample powderization gives rise to a significant portion of martensitic traces inside the high-T austenitic region, and that these traces can be subsequently removed by annealing. 
\end{abstract}


\maketitle

\section{Introduction}
The discovery of a martensitic phase in Ni$_2$MnGa below the ferromagnetic (FM) Curie point by Webster {\it et al.} in 1984,\cite{Webster84} has triggered an extensive experimental and theoretical activity on Heusler  and related alloys over the last three decades.\cite{Vasilev03,Entel06} Besides their fundamental interest, the complex interplay of structural, magnetic and electronic degrees of freedom in these compounds gives rise to technologically functional properties such as magnetic shape memory,\cite{Ullakko96} magnetocaloric,\cite{Hu00} as well as magnetoresistance effects.\cite{Biswas05}

Ni$_2$MnGa is a FM Heusler alloy with Curie temperature T$_\textrm{c}\!\simeq$ 380~K.\cite{Webster84} 
The magnetic moment is mainly localized on Mn sites, while on Ni sites the magnetic moment is much smaller 
(a tenth of the Mn moment).\cite{Webster84} Still, the conduction electrons from Ni seem to play an important role in mediating the Mn-Mn interactions and the ferromagnetic ordering in Ni$_2$MnGa.\cite{Kubler83}

Upon cooling below T$_\textrm{c}$, Ni$_2$MnGa shows two thermally driven structural transitions, one from the high temperature austenitic to the so-called premartensitic (PM) phase at T$_{\textrm {PM}}\simeq$ 260 K, and another from the PM to the martensitic phase at T$_{\textrm {M}}\simeq$ 200 K.\cite{Webster84} In the austenitic phase, Ni$_2$MnGa has the fcc L$2_1$ crystal structure with \textit{a}=5.825 \AA~ and space group (SG) \emph{Fm}$\overline{3}$\emph{m} (No.~225).\cite{Webster84} 
The PM transition proceeds via a pronounced softening in the $\left[\zeta \zeta 0\right]$ TA$_2$ phonon branch at $\zeta\!=1/3$, as observed by inelastic neutron scattering (INS) measurements reported by Zheludev {\it et al.}\cite{Zheludev95} The freezing of the displacements associated with this softening gives rise to a distortion of the austenitic structure with the propagation vector of the soft mode.\cite{Zheludev95} This softening, which has also been observed in other shape memory alloys with similar structure,\cite{Shapiro84,Shapiro91} can be ascribed to the interplay of strong electron-phonon coupling and Fermi surface (FS) nesting\cite{Zhao89,Zhao92,Lee02,Bungaro03,Zayak03,Bungaro03,Zayak03} 
(see also the phenomenological model by Planes {\it et al.}\cite{Planes97} and the recent first principle calculations by Uijttewaal {\it et al.}\cite{Uijttewaal09}). 
Besides the structural modifications,\cite{Zheludev95,Zheludev96,Worgull96,Stenger98} the PM transition is also accompanied by small field-dependent magnetization changes at T$_{\textrm{PM}}$.\cite{Zuo98}

Turning to the martensitic phase below T$_{\textrm {M}}\simeq$ 200 K,\cite{Webster84} 
its crystal structure and space group remain under debate. Initially, the martensitic phase was described as a tetragonal distortion of the parent phase with \textit{a}=5.920 \AA ~ and \textit{c}=5.566 \AA, which in addition has a superstructure with long periodicity along the \textit{c} axis.\cite{Webster84} 
Later, Martynov {\it et al.}\cite{Martynov92} reported that the superstructure can be described as a periodic shuffling of the (100) planes along the [$\bar{1}$00] direction with a periodicity equal to 5 atomic layers (five-fold modulation, 5M). Neutron scattering experiments by Zheludev {\it et al.},\cite{Zheludev96} showed that the 5M picture should be actually described as an incommensurate modulation with wave vector $(0.43,0.43,0)$, which amounts to nearly 5 interplanar distances, while  
Brown {\it et al.}\cite{Brown02} inferred that the martensitic phase is 7M with orthorhombic symmetry and SG \emph{Pnnm} (No.~58). Orthorhombic symmetry was also reported by Righi {\it et al.},\cite{Righi06} based upon X-ray powder diffraction experiments, but with a modulation vector \textbf{q}=0.43$\textbf{c}^*$, as reported in Ref.~\onlinecite{Zheludev96}. 

Besides the exact crystal structure of the martensitic phase, which is sensitive to stoichiometry,\cite{Vasilev03,Entel06} the microscopic origin of the martensitic transition is also debated, the main proposals being the band Jahn-Teller mechanism\cite{Fujii89,Brown99,Ayuela02,Barman05} on one hand, and strong electron phonon-coupling and FS nesting\cite{Zheludev95,Zheludev96} on the other. The latter has been supported by extended \textit{ab initio} calculations which
have succeeded into reproducing several experimental findings.\cite{Lee02,Bungaro03,Zayak03} Recent neutron scattering experiments by Shapiro {\it et al.},\cite{Shapiro07} reported well defined phason excitations which were associated to the charge density wave (CDW) resulting from FS nesting.\cite{Bungaro03}
Furthermore, ultraviolet-photoemission measurements have shown the formation of a pseudogap 0.3 eV below the Fermi energy at T$_{\textrm {PM}}$,\cite{Opeil08,Souza12} which has also been attributed to CDW due to the FS nesting.\cite{Ayuela02}
A third proposal for the origin of the martensitic transition is the concept of adaptive modulations,\cite{Khachaturyan91} 
which was recently applied to Ni$_2$MnGa.\cite{Kaufmann10} In this scenario, the stabilization of the martensitic phase is not of electronic origin,
instead modulated martensites are formed by nanotwinned variants of the tetragonal phase.

Here we present a zero-field static and dynamic $^{55}$Mn NMR study of bulk and powder samples of Ni$_2$MnGa alloy,
complemented by magnetization and X-ray powder diffraction experiments. 
The NMR experiments provide local access to the above sequence of structural phase transitions from the high-T austenitic
to the premartensitic and finally to the low-T martensitic phase. In addition, a careful study of the so-called {\it rf}-enhancement 
factors allows to probe the stiffness and local anisotropy for each separate magnetic environment. In this way, we are able to differentiate the signals from the
austenitic and martensitic components and follow their evolution with temperature. Our measurements on bulk and powdered samples demonstrate strong dependence on sample preparation. Most notably, we find that powderization leads to the formation of martensitic traces already at high temperatures, which can subsequently be removed by annealing.
Our article is organized as follows. In Sec.~\ref{expdetails} we provide the experimental details.
In Sec.~\ref{xrd} we present the X-ray powder diffraction experiments. In Sec.~\ref{Magnetization} we present our magnetization measurements and
in Sec.~\ref{NMR} the NMR results. Finally, a brief summary of our results is given in Sec.~\ref{sec:disc}.

\section{Experimental details}\label{expdetails}
Polycrystalline samples of Ni$_2$MnGa were prepared by the repeated arc-melting of stoichiometric quantities of the starting elements in an arc discharge furnace. The ingot was annealed for homogenization at 800$^\circ$C for two weeks and later was cut in two pieces. Hereafter, we will call the first piece of the ingot ``sample-I''. The second piece of the ingot was crushed into powder and will be referred to as ``sample-II'' in the following. Both samples were studied by magnetization and NMR measurements, while sample-II was also studied by X-ray powder diffraction at room temperature. Upon completing these experiments sample-II was sealed under low argon pressure in a quartz ampule, and annealed for 4 days at 600$^\circ$C. The ampule was subsequently quenched in iced water. This procedure was repeated one more time. We will refer to the annealed sample produced by the aforementioned process as `sample-III''. In this particular sample we have performed, apart from the magnetization and NMR experiments, X-ray powder diffraction experiments as a function of temperature. 

\begin{figure}[!t] 
\centering
\includegraphics[width=0.49\textwidth] {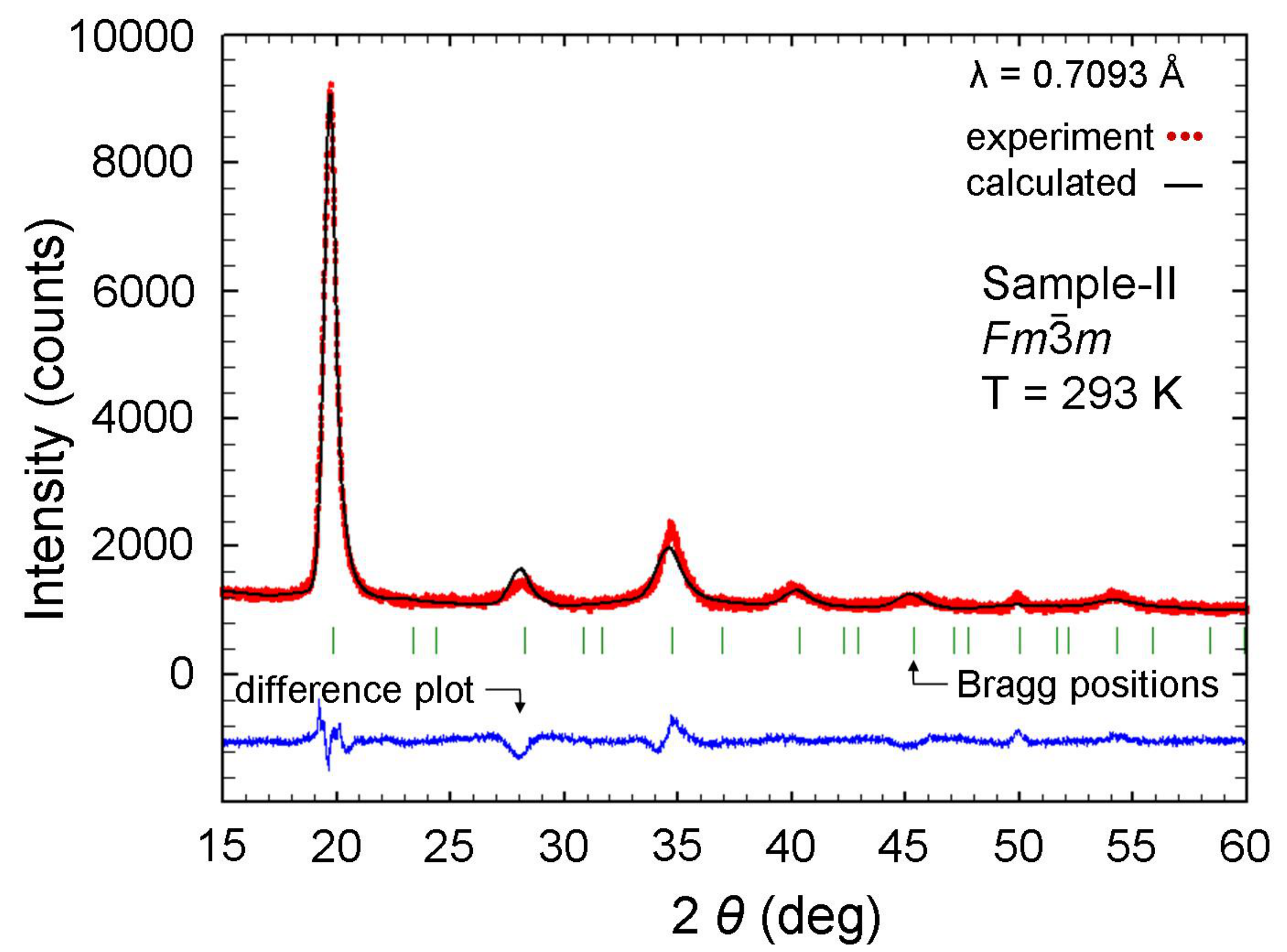} 
\caption{(Color online) The  X-ray powder diffraction pattern of sample-II at 293~K.}
\label{sampleIIXRD}
\end{figure}

The X-ray powder diffraction experiments were performed on a STOE Stadi P powder diffractometer 
with Mo K$_{\alpha1}$ radiation. Sample-II was studied at 293~K and sample-III between 140-293 K. The diffractometer is equipped with a curved Ge(111) monochromator and a 6$^\circ$-linear position sensitive detector. Sample-II was measured in transmission geometry as flat sample with a thin powder layer glued onto a polyacetate film. Sample-III was filled in a capillary, which was afterwards sealed and measured in Debye-Scherrer mode with a step size of $0.01^\circ$ and 100 s/step in the range 15$^\circ\leq 2\theta\leq 60^\circ$. The data were evaluated by the Rietveld method~\cite{Rietveld69} with Fullprof in the WinPlotR program package.\cite{Roisnel01} For T$\geq$170~K mainly the austenitic phase, SG \emph{Fm}$\overline{3}$\emph{m}~\cite{Soltys75}, was used as structure model, while for T$\leq$200~K the 7M in-phase model with SG \emph{Pnnm}~\cite{Brown08} was taken as second phase. For the refinements, a Thompson-Cox-Hastings
pseudo-Voigt profile function was selected.\cite{Thompson87} As refinable parameters background, scale factor, half width, Caglioti variables (U, V, W), lattice parameters, asymmetries and the overall temperature factor B$_{ov}$ were allowed. For cooling, an Oxford Cryosystems 700 series equipment was used to cycle the sample within a temperature range of 140-293~K. At every temperature, the sample was equilibrated for 1 h before starting the measurement.

The magnetization measurements were performed with a SQUID magnetometer (MPMS, Quantum Design) in the temperature range of 2-400~K and for applied fields up to 5 T. The temperature dependence of the magnetization was measured in zero field cooling (ZFC) and in field cooling (FC) modes. In the ZFC mode, the sample was first cooled to 2 K in zero field, then a magnetic field was applied and the data were collected while heating. In the field cooled (FC) measurements, the magnetic field was applied above the transition temperature to the ferromagnetic state and the data were taken during cooling.

The NMR experiments were performed with a Redstone-TECMAG spectrometer (10-500 MHz),
which is interfaced with a power level meter and the NMR probe-head (NMR-Service). The latter is 
equipped with computer controlled step motors which allow fully automated tuning and matching of the 
tank circuit, ensuring minimal reflected {\it rf} signal over a very broad frequency range. The setup is 
supplemented by a Janis cryostat and a Lakeshore temperature controller which allow measurements 
in the range 1.5-300~K.  The $^{55}$Mn NMR signals were obtained by a $0.8\mu$s-$\tau$-$0.8\mu$s 
spin-echo pulse sequence where the separation between the {\it rf} pulses was $\tau=5\mu$s. 
It is well known that, in ferromagnets, the applied {\it rf} field $H_1$ and the induced NMR signals 
are enhanced by a factor known as ``{\it rf} enhancement factor $\eta$'' (see Sec.~\ref{enhancement}).\cite{Gossard59} 
The NMR spectra presented here are corrected for the {\it rf} enhancement factor and thus the relative 
intensities are proportional to the number of resonating nuclei at each different frequency at time $2\tau$. 
The protocol followed here is similar to Refs.~\onlinecite{Panissod00,Panissod02}. The spin-lattice relaxation time
was measured at the peak of the austenitic and martensitic spectra, by applying a saturation recovery technique and by fitting with a single exponential recovery law.

\section{Experimental results}\label{expresults}
\subsection{X-ray powder diffraction}\label{xrd}
{\it Sample-II ---} Figure \ref{sampleIIXRD} shows the room temperature X-ray pattern of sample-II. 
We find broadened reflections with different shapes, a high signal-to-noise ratio and
an isotropic peak shift to higher angles. As an additional feature, a peak asymmetry located to higher 
angles is observed, especially for the 220, 442 and 444 reflections. Two possible scenarios may lead to 
this behavior: (i) stress/strain effects, (ii) a second phase or a distortion of the observed cubic phase, 
or a concomitant overlay of both (i) and (ii). Especially the second scenario, involving a second phase, 
probably a martensitic, seems to be consistent with the NMR results of sample-II. The lattice parameter \emph{a} 
for the \emph{Fm}$\overline{3}$\emph{m} structure model used for refinement is determined to 5.8161(9) {\AA} 
with an unit cell volume \emph{V}=196.7(1) {\AA}$^{3}$. 

\begin{figure}[!t] 
\centering
\includegraphics[width=0.49\textwidth] {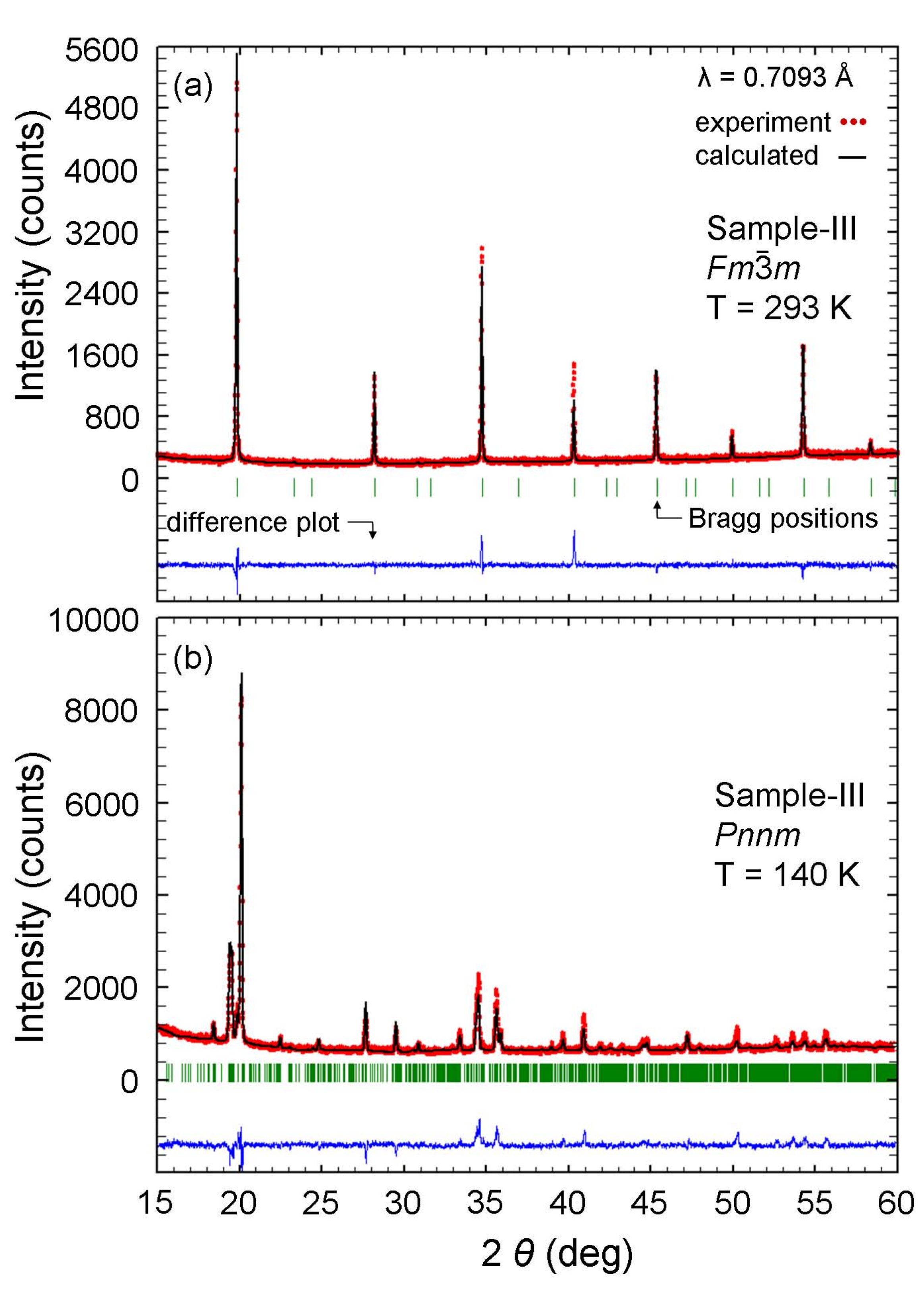} 
\caption{(Color online) The  X-ray powder diffraction patterns of sample-III in (a) the austenitic
(\emph{Fm}$\overline{3}$\emph{m}) phase at 293~K and (b) in the martensitic (\emph{Pnnm}) phase at 140 K.}
\label{XRD}
\end{figure}

\begin{table}[!b]
\caption{(Color online) Lattice parameters and phase contents (p.c.) of
the temperature-dependent XRD experiments on sample-II. The
temperature regime is shown from cooling to heating.}\label{tab:xrd}
\begin{ruledtabular}
\begin{tabular}{*7c}
 \emph{T}(K) & SG & \emph{a}({\AA}) & \emph{b}({\AA}) & \emph{c}({\AA}) & \emph{V}({\AA}$^{3}$) & p.c.\\ 
\hline
   293 & 225 & 5.8210(1) &&& 197.24(1) & \\
   230 & 225 & 5.8139(1) &&& 196.52(1) & \\
   200$^a$ & 225 & 5.8124(1) &&& 196.364(2) & \\
   180$^a$ & 225 & 5.8116(1) &&& 196.29(1) & \\
   170$^b$ & 225 & 5.8092(2) &&& 196.05(3) & 65\%\\
    			 & 58 & 4.2051(7) & 29.275(4) & 5.5808(7) & 687.0(3) & 35\%\\
   140$^b$ & 58 & 4.2044(4) & 29.261(2) & 5.5672(4) & 684.9(2) & \\
   170$^b$ & 58 & 4.1969(17) & 29.224(9) & 5.5760(15) & 683.9(7) & 31\%\\
    			 & 225 & 5.8045(2) & & & 195.98(2) & 69\%\\
   180$^a$ & 225 & 5.8104(2) & & & 196.17(3) & \\
   200$^a$ & 225 & 5.8121(2) & & & 196.34(2) & \\
   230 & 225 & 5.8125(2) & & & 196.37(2) & \\
   293 & 225 & 5.8176(1) & & & 196.90(1) & \\
\hline
\multicolumn{7}{l}{$^\text{a}$ \emph{Pnnm} phase is already visible, but not refinable. }\\
\multicolumn{7}{l}{~~All parameters were set manually.}\\
\multicolumn{7}{l}{$^\text{b}$ B$_{ov}$ set to zero, otherwise becomes negative}\\
\multicolumn{7}{l}{~~related to stress/strain or other texture effects.}
\end{tabular}
\end{ruledtabular}
\end{table}

{\it Sample-III ---} Low temperature X-ray powder diffraction was measured on a powder of sample-III. 
The Rietveld refinements were performed with focus on phase transition, phase content and lattice parameters. 
All results are presented in Table 1. The X-ray patterns of sample-III at 293 K and 140 K are displayed in 
Fig.~\ref{XRD}. The lattice parameter \emph{a} of the cubic phase (\emph{Fm}$\bar{3}$\emph{m}, Fig.~\ref{XRD}(a))
changes upon cooling from 5.8210(1)~{\AA} at 293~K to 5.8092(2)~{\AA} at 170~K. During warming up, a small 
hysteresis is observed and the lattice parameter \emph{a} does not reach its initial value at 293~K.
The onset of the martensitic phase (\emph{Pnnm}, Fig.~\ref{XRD}(b)) takes place at 200~K and finishes at 140~K, with lattice parameters 
of \emph{a}=4.2044(4)~{\AA}, \emph{b}=29.261(2)~{\AA} and \emph{c}=5.5672(4)~{\AA}. The evolution of the phase transition 
is similar for the warming branch where the transition is almost completed at 200~K. Nevertheless, it has to be 
noted that a martensitic phase may be present at higher temperatures in both temperature cycles but it is 
neither refinable nor the reflections are well-defined. This fact is easily demonstrated on the theoretically most 
intense 172 reflection which is visible with a signal-to-noise ratio of $\approx$ 1.1. Additionally a complex microstructure 
of the investigated Heusler compound cannot be excluded. Texturing or stress/strain effects (Fig.~\ref{XRD}(a)) may be the reasons 
for residual non-refinable intensities. 

\subsection{Magnetization measurements}\label{Magnetization}
\begin{figure}[t] 
\centering
\includegraphics[width=0.49\textwidth] {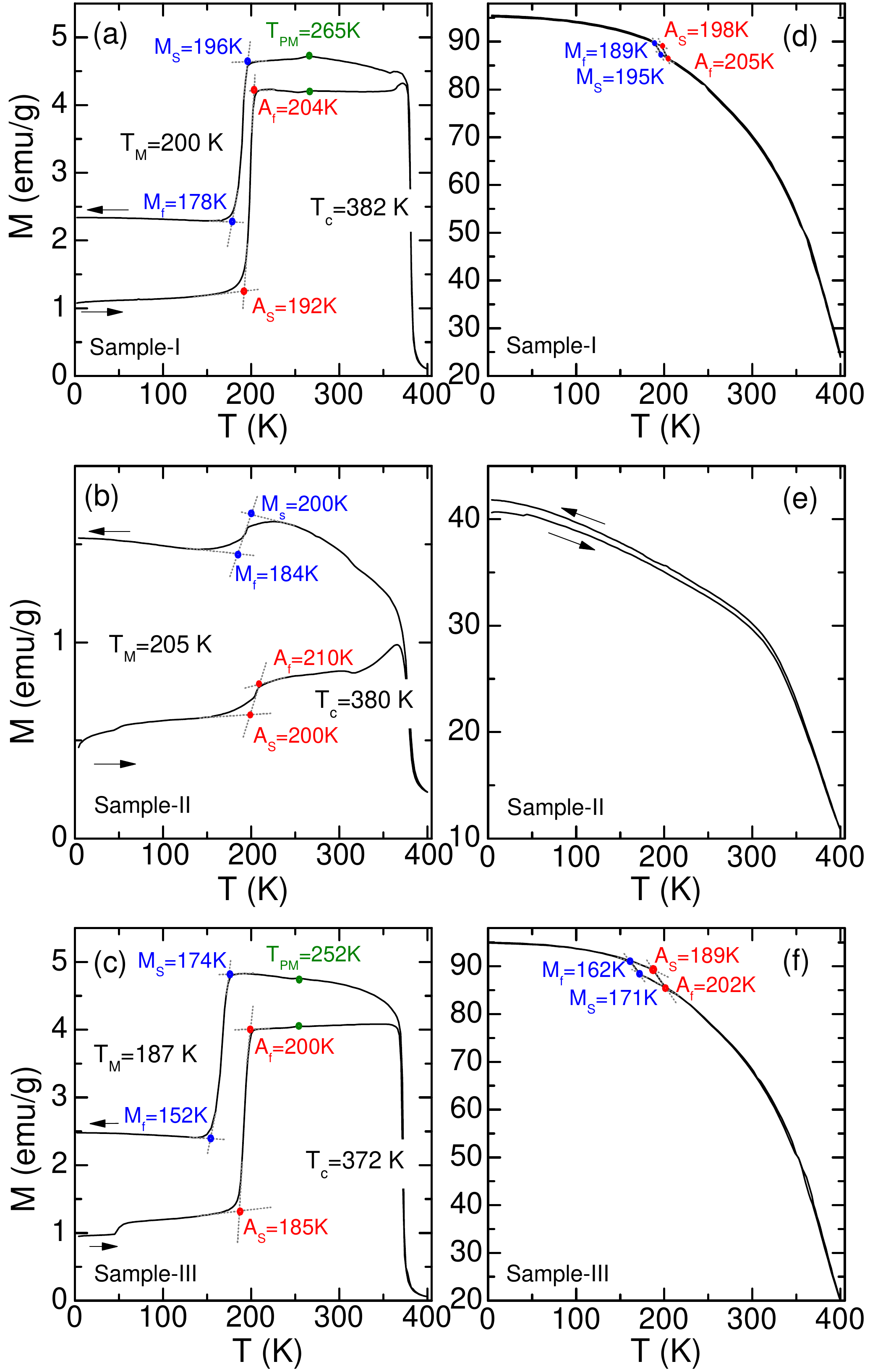} 
\caption{(Color online) ZFC and FC magnetization loops measured at 100 Oe for samples I, II and III in (a)-(c) respectively. The 
corresponding high field (H$=4$ T) data are presented in (d)-(f).}
\label{SQUID}
\end{figure}

Figures~\ref{SQUID}(a)-(c) show the ZFC/FC magnetization curves at 100 Oe for the three samples investigated in this work, while Figs.~\ref{SQUID}(d)-(f) show the corresponding data at 4~T.

{\it Sample-I ---}
From the low field magnetic measurements in sample-I (see Fig.~\ref{SQUID}(a)) we find a jump 
at T$_\textrm{c}\!=$382~K (taken as the minimum point of $dM/dT$) which corresponds to the transition from the PM to the FM phase. Right below T$_c$ we observe the so-called Hopkinson peak~\cite{Wang09}. At lower temperatures we observe a small bump at T$_{\text{PM}}\!=$265 K which is attributed to the premartensitic transition.\cite{Zuo98} Finally, the large drop at T$_{\text{M}}\!=$~200 K is due to the onset of the martensitic phase which, as expected, has higher magnetic anisotropy compared to the cubic austenitic phase.\cite{Webster84} The transition temperature T$_\textrm {M}$ is estimated by T$_{\text{M}}\!=\!\left( \textrm{M}_\textrm{s}+\textrm {A}_\textrm{f} \right)/2$, where the martensitic (austenitic) transformation temperatures upon cooling (warming) M$_{\text{s}}$ (A$_{\text{s}}$), M$_{\text{f}}$ (A$_{\text{f}}$) are marked in Fig.~\ref{SQUID}. The magnetic measurements performed at H= 4~T, which is higher than the saturating field at all temperatures, show a small increase of the magnetization by 1.8 emu/g (0.078 $\mu_B$ per formula unit) at the onset of the martensitic transition. This increase is in agreement with {\it ab initio} electronic structure calculations by Opeil {\it et al.}\cite{Opeil08} which show a spectral weight transfer from the spin-down to the spin-up channel at the martensitic transition. Our NMR experiments presented in Sec.~\ref{NMR} below give an independent confirmation of this fact.

\begin{figure}[t] 
\centering
\includegraphics[width=0.4\textwidth] {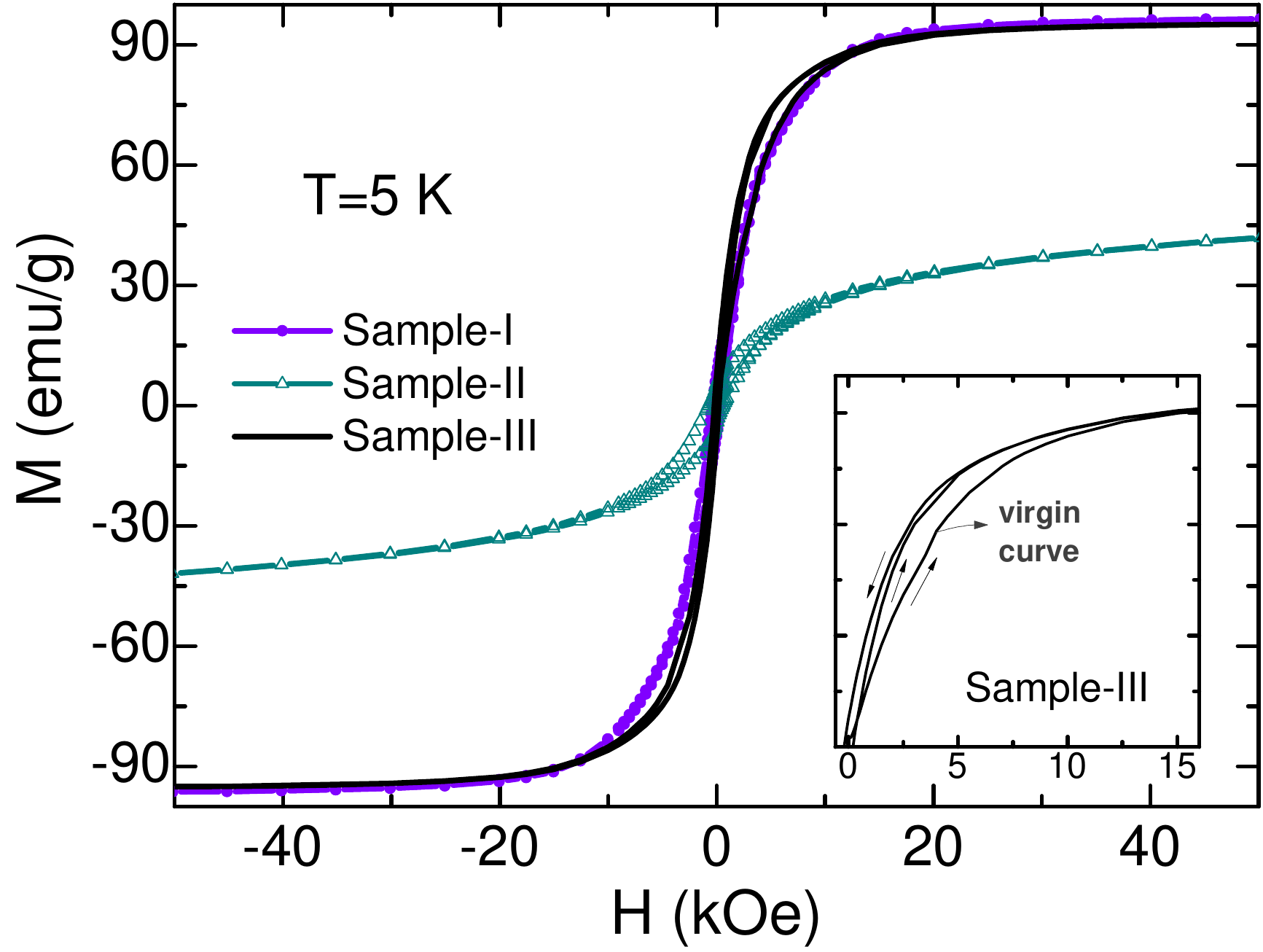} 
\caption{(Color online) Magnetization loops at 5 K for samples I, II and III. The inset offers a magnified view of the M-H loop for sample-III.}
\label{SQUIDloop}
\end{figure}

{\it Sample-II ---}
Here the transition to the ferromagnetic state occurs at T$_{\text{c}}\!=$380~K (Fig.~\ref{SQUID}(b)), 
which is very close to the value found in sample-I. However, the magnetization of sample-II is considerably lower and the jump at the martensitic transition is much weaker. This behavior shows that after powderizing sample-I, the magnetic anisotropy is higher in both the austenitic and martensitic phases of sample-II. This is also evident in the magnetization loops (Fig.~\ref{SQUIDloop}), where harder magnetic behavior is observed, but also in the NMR experiments presented below in Sec.~\ref{enhancement}. The hard magnetic behavior is further reflected in the high field M-T data in Fig.~\ref{SQUID}(e)), where we also note that no anomaly is observed at T$_{\text{M}}$ as in the low field data of Fig.~\ref{SQUIDloop}(b). This behavior can be explained based on the NMR data in Sec.~\ref{NMR}, which show the presence of martensitic precursors in sample-II already at room temperature (the highest-T in our NMR experiments).

{\it Sample-III ---} 
Here we find that the annealing treatment (Sec.~\ref{expdetails}) resulted in recovering the magnetization jump at the onset of the martensitic transition, which can be observed both in low and high magnetic fields (Figs.~\ref{SQUID}(c), (f)). This behavior shows that by annealing we have lowered the magnetic anisotropy and have eliminated the aforementioned martensitic precursors. On the other hand, the transition temperatures T$_{\text{c}}$, T$_{\text{PM}}$ and T$_{\text{M}}$ are lower in sample-III compared to samples-I and II, and the Hopkinson peak disappears from the low-H data. The latter implies lower magnetic anisotropy in the austenitic phase of this particular sample, which as we will see in Sec.~\ref{NMR} is in agreement with the NMR results. Furthermore, a small downturn is observed at 55 K in the low field ZFC data. Similar behavior has been observed in Ni-Mn-Sn~\cite{Li07}, Ni-Mn-Sb~\cite{Khan07}, and Ni-Mn-In~\cite{Wang08}, and was attributed to coexisting FM and AFM phases, which gives rise to an exchange bias effect that manifests in
shifted hysteresis loops. Here, the presence of this anomaly is not accompanied by a shift in the magnetization loops and may be associated with the presence of austenitic traces down to low temperatures according to our zero field NMR experiments presented in Sec.~\ref{enhancement}. We anticipate that the coexistence of martensitic and austenitic phases, which have distinct magnitude and T-dependence of magnetocrystalline anisotropy, can be responsible for the small downturn observed in Fig.~\ref{SQUID}(c).

\begin{figure*}[!t] 
\centering
\includegraphics[width=0.98\textwidth]{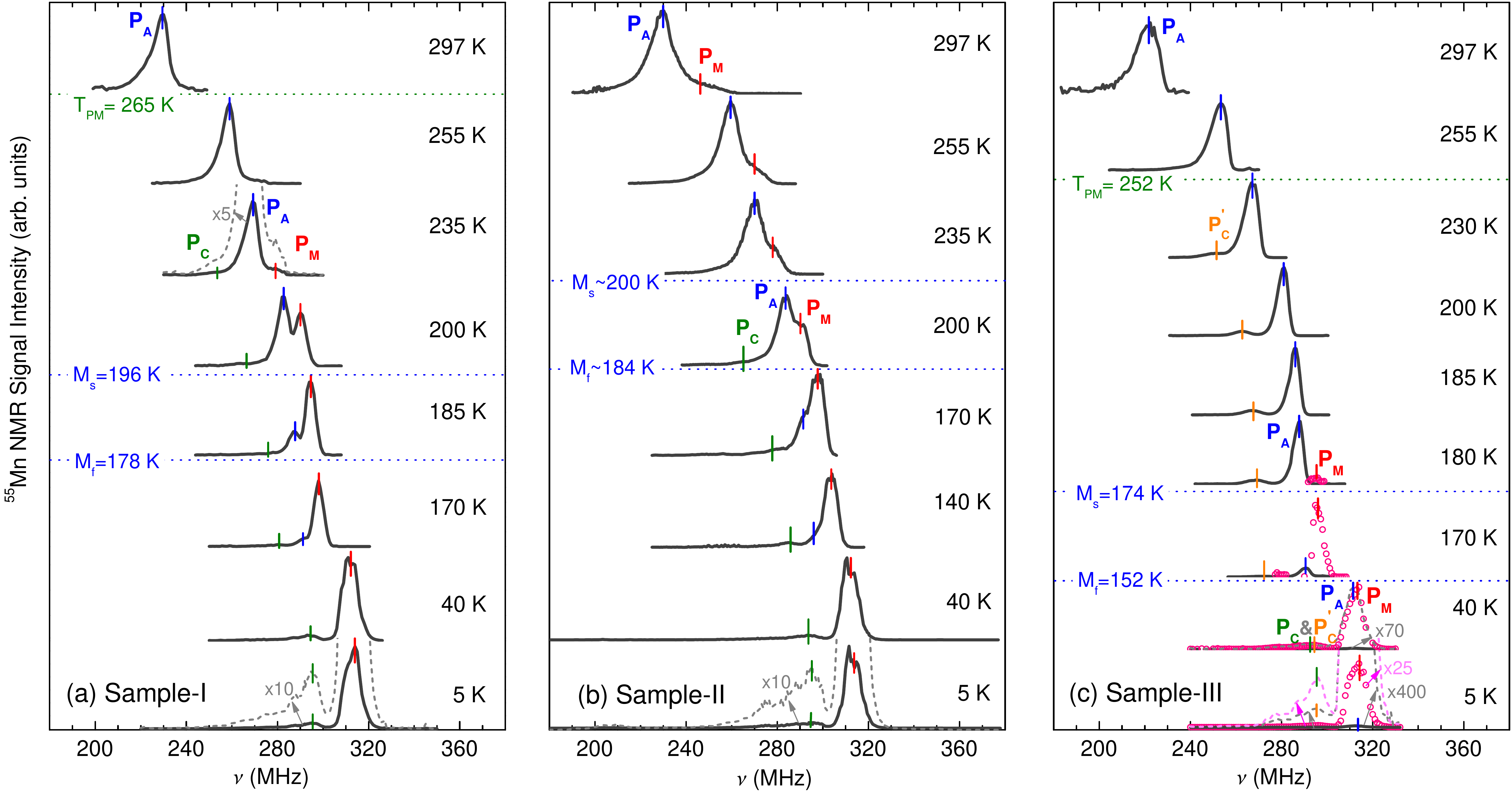} 
\caption{(Color online) Zero-field $^{55}$Mn NMR spectra of polycrystalline Ni$_2$MnGa samples acquired upon cooling. (a) Sample-I, 
(b) sample-II and (c) sample-III at various temperatures. With dotted lines we provide
enlarged views at some characteristic temperatures. The transformation temperatures obtained 
from the magnetization measurements are also indicated.}
\label{allLines}
\end{figure*}

{\it Magnetization loops ---} 
From the magnetization loops collected at T=5~K (Fig.~\ref{SQUIDloop}), we find that the magnetic moment of sample-I at 
saturation is M=96 emu/g, as in other works,\cite{Webster84} and the anisotropy field (obtained from the anomaly in $d^2 M/dH^2$) 
is H$_A$=~3.5~kOe, which is a typical value for the stoichiometric Ni$_2$MnGa.\cite{Tickle99,Chu00,Albertini01} In contrast, in 
sample-II we find that an applied field of 5 T is not sufficient to saturate the magnetization, which indicates that powderizing 
the sample induces large internal stresses. This is further supported by the X-ray powder diffraction data (Sec.~\ref{xrd}) as well as by the NMR {\it rf}-enhancement experiments presented below in 
Sec.~\ref{enhancement}. Turning to sample-III, we find that the annealing process has recovered the soft magnetic behavior and the 
high value of the saturation magnetization. Please note that since no correction for the demagnetizing field has been included in 
the data of Fig.~\ref{SQUIDloop}, a direct quantitative comparison of the anisotropy field values between sample-I (bulk) and sample-III (powder) is not possible.

A noticeable feature of sample-III is that the virgin magnetization curve lies below the returning loops, see inset of Fig.~\ref{SQUIDloop}. Similar behavior has been observed in many systems such as the off-stoichiometric Ni-Mn-In~\cite{Sharma07}, Ni-Mn-Sn~\cite{Chatterjee08}, Ni-Co-Mn-Sn~\cite{Banerjee11}, and Ni-Co-Mn-Sb~\cite{Nayak10}, where it is attributed to the competition between coexisting FM and AFM phases across a disorder influenced first-order transition. 
Here the above feature is attributed to the presence of austenitic traces well inside the martensitic phase (as unveiled by the zero field NMR experiments presented in Sec.~\ref{enhancement}), since the applied magnetic field generally enhances the fraction of the austenitic phase and thus the bulk magnetization at low fields. Another possible explanation is the magnetic field induced rearrangement 
of the martensitic variants, besides the rotation of magnetization.\cite{Heczko02,Jing-Min06,Okamoto05}

\subsection{Nuclear Magnetic Resonance}\label{NMR}
\subsubsection{$^{55}$Mn NMR lineshapes}\label{lines}

The zero field $^{55}$Mn NMR spectra were measured in the temperature range 5-297 K upon cooling for all samples and the results are presented in Fig.~\ref{allLines}.

{\it Sample-I ---}
At high temperatures we observe one peak which corresponds to the austenitic phase and is indicated 
as P$_{\textrm{A}}$ in Fig.~\ref{allLines}(a). The presence of one line in the NMR spectrum is indeed 
expected since in the austenitic phase of Ni$_2$MnGa we have one Mn site with octahedral site symmetry and 
thus the quadrupolar splitting is zero. However a small asymmetry of the line towards low frequencies is observed. 
This could be explained either by the presence of a small disorder in the system, or by small deviations from the 
cubic structure as we are close to the transition into the premartensitic phase. 

At T$\sim$ 235 K a new peak, P$_{\textrm{M}}$, appears at higher frequencies compared to P$_{\textrm{A}}$, 
whose intensity grows by decreasing temperature. At the same time the intensity of the austenitic peak 
P$_{\textrm{A}}$ decreases until it disappears from the spectrum at around 140 K. The peak P$_{\textrm{M}}$, 
which was first reported in Ref. [\onlinecite{Ooiwa92}], dominates the NMR spectrum at low temperatures and thus 
originates from the manganese sites in the martensitic phase of the sample. This is further supported by the {\it rf}-enhancement 
experiments presented in Sec.~\ref{enhancement}, which in addition shed light on the local magnetic anisotropies of austenitic 
and martensitic phases. It is interesting to note that the martensitic peak P$_{\textrm{M}}$ is shifted by 8 MHz (0.762 T), 
compared to the austenitic peak P$_{\textrm{A}}$. For a typical value of the hyperfine coupling constant A=10 T$/\mu_B$,\cite{Akai90} we find that the enhancement of the hyperfine field corresponds to an increase of the magnetic moment by 0.076 $\mu_B$ in 
the martensitic phase. This value is in nice agreement with our high field magnetization data (Sec.~\ref{Magnetization}) and with electronic structure calculations which
have shown that the martensitic transformation in Ni$_2$MnGa is accompanied by a spectral weight transfer from spin-down to spin-up electrons.\cite{Opeil08}

Apart from the two main peaks P$_{\textrm{A}}$ and P$_{\textrm{M}}$, we also note that below 235 K the low frequency tail of the NMR spectrum begins to develop into a small peak, which hereafter we will indicate as P$_{\textrm{C}}$. The peak P$_{\textrm{C}}$ is observed down to low temperatures where its fine structure is unveiled. Specifically, apart from P$_{\textrm{C}}$, we can distinguish three smaller equidistant peaks at lower frequencies with an average frequency shift among them of approximately 8 MHz.

\begin{figure}[t] 
\centering
\includegraphics[width=0.49\textwidth]{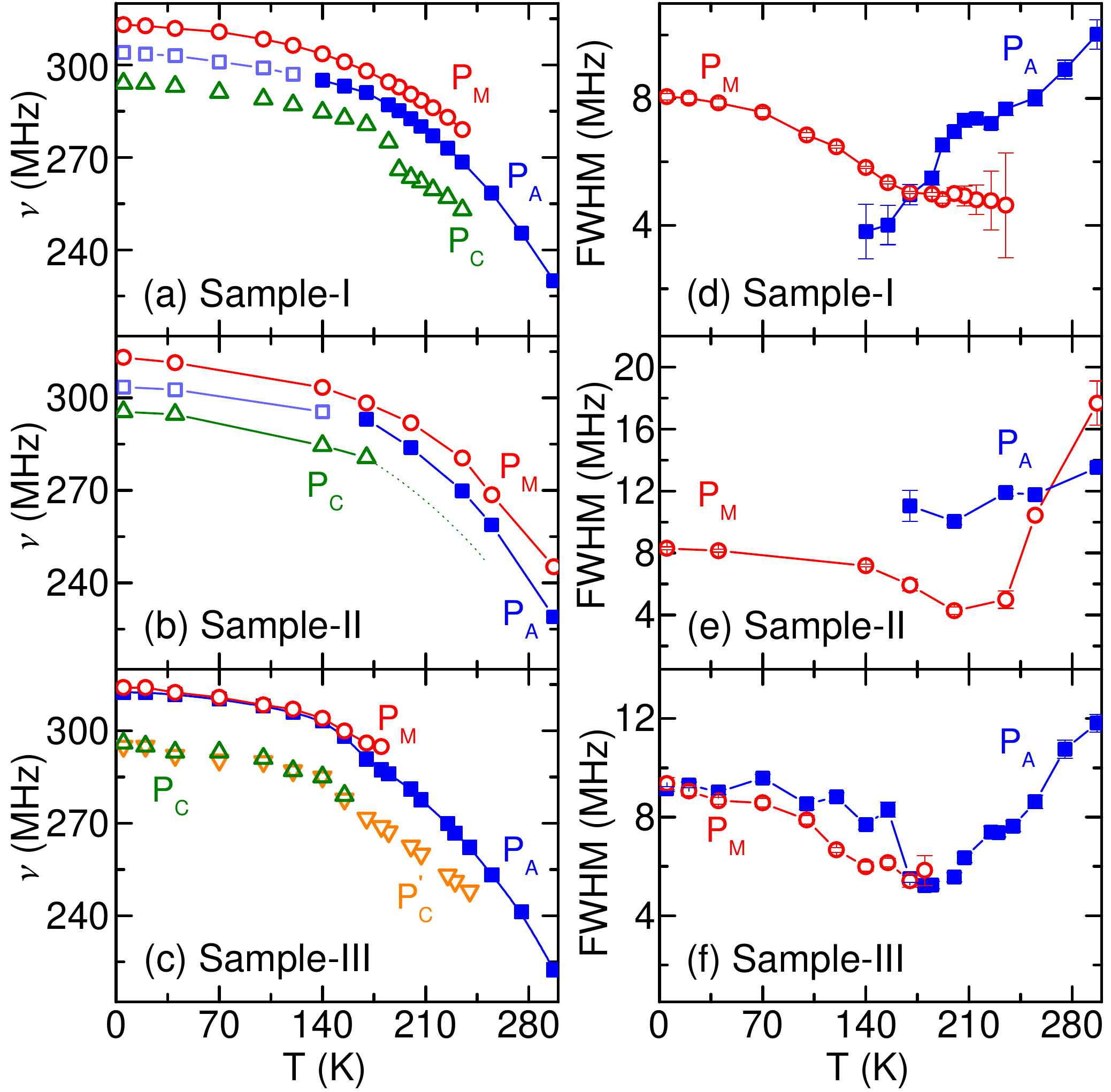} 
\caption{(Color online) Temperature dependence of the $^{55}$Mn NMR resonance frequency for the main peaks observed in (a) sample-I, (b) sample-II and (c) sample-III. With open squares in (a) and (b) we give the resonance frequency of austenitic remnants found at low temperatures from the {\it rf}-enhancement experiments presented in Sec.~\ref{enhancement}. The temperature dependence of the full width at half maximum (FWHM) for the austenitic peal P$_A$ and martensitic peak P$_M$ for (d) sample-I, (e) sample-II and (f) sample-III. Lines are guides to the eye.}
\label{FvsTandFWHMall}
\end{figure}

{\it Sample-II ---}
The NMR lineshapes of sample-II (Fig.~\ref{allLines}(b)) show similarities but also a few differences compared to the corresponding spectra of the parent compound. The main difference to sample-I is that upon powderizing we have created traces of the martensitic phase (peak P$_{\textrm{M}}$) already from high temperatures. The high frequency peak indicated as P$_{\textrm{M}}$ in Fig.~\ref{allLines}(b) is readily attributed to the martensitic phase, for two main reasons. At first, its resonance frequency is smoothly connected to the martensitic peak at lower temperatures, as can be seen in Fig.~\ref{FvsTandFWHMall}(b). The second argument comes from the {\it rf}-enhancement experiments presented in Sec.~\ref{enhancement}. The {\it rf} power required to excite the nuclei contributing to peak P$_{\textrm{M}}$ at high temperatures is the same as the one applied deep inside the martensitic phase and it is clearly distinct from the typical values found in the austenitic phase of sample-I. These findings imply that the stresses induced upon the sample preparation have created traces of the martensitic phase already at room temperature, which is also in line with our X-ray powder diffraction data. The observation of martensitic precursors explains the magnetization measurements of sample-II, i.e. the absence of magnetization jumps at T$_M$ in the high field measurements (Fig.~\ref{SQUID}(e)) and the strong magnetic anisotropy of the high-T phase (Figs.~\ref{SQUID}(e), \ref{SQUIDloop}). 

Besides the appearance of peak P$_{\textrm{M}}$ at high T, the behavior of sample-II is similar to sample-I. The small peak P$_{\textrm{C}}$ is also present here, as well as the austenitic peak P$_{\textrm{A}}$. Peaks P$_{\textrm{A}}$ and P$_{\textrm{C}}$ have similar resonance frequency and temperature dependence (Fig.~\ref{FvsTandFWHMall}(b)) as the corresponding peaks in sample-I (Fig.~\ref{FvsTandFWHMall}(a)). We note though that the full width at half maximum (FWHM) of peak P$_{\textrm{A}}$ is larger in sample-II compared to the bulk parent compound, indicating enhanced inhomogeneities (magnetic or/and structural) in the austenitic phase of the powdered sample. 

{\it Sample-III ---}
After annealing sample-II, we find that the high temperature (T$>$ 200~K) signal from the martensitic phase P$_{\textrm{M}}$ 
disappears. Instead, the martensitic peak P$_{\textrm{M}}$ shows up at
intermediate temperatures as found in sample-I. This indicates that by annealing we have released the internal strains and eliminated the martensitic phase grown in sample-II upon its powderization process. This is in agreement with our X-ray powder diffraction data and is further supported by the smaller FWHM of the austenitic peak P$_{\textrm{A}}$ in sample-III compared to sample-II, which are 
plotted in Fig.~\ref{FvsTandFWHMall}(b) and (c) respectively. The resonance frequency for the austenitic peak P$_{\textrm{A}}$ at high temperatures, as well as for the martensitic peak P$_{\textrm{M}}$ for sample-III (Fig.~\ref{FvsTandFWHMall}(c)), are slightly smaller than the corresponding values of the other samples. The intensity of the austenitic peak P$_{\textrm{A}}$ decreases on cooling, while that of the martensitic peak P$_{\textrm{M}}$ increases, as found in the other samples. 

Sample-III shows austenitic remnants down to 5 K, as can be seen in the lineshape measurements of Fig.~\ref{allLines}(c). At low temperatures the NMR signals from austenitic and martensitic phases overlap in frequency, but the distinction between the two is still possible due to their very different {\it rf}-enhancement factors, see Sec.~\ref{enhancement}. We note here that the amount of austenitic remnants is quite low (less than 0.1\%) and thus not possible to detect with the X-ray powder diffraction data. Apart from peaks P$_{\textrm{A}}$ and P$_{\textrm{M}}$ we also notice that the low frequency tail of the lines, present at high temperatures, transforms into a small peak P$_{\textrm{C}}'$ around 230 K (Fig.~\ref{FvsTandFWHMall}(c)), in a similar way with the other two samples. The peak P$_{\textrm{C}}'$ is observed down to low temperatures where it overlaps with the much stronger martensitic peak P$_{\textrm{C}}$. Furthermore, as has been found in the other two samples, apart from P$_{\textrm{C}}'$ and P$_{\textrm{C}}$, some weaker equidistant peaks are again observed at lower frequencies. 

\subsubsection{rf enhancement}\label{enhancement}
In ferromagnets the strong hyperfine field, $\textrm{H}_\textrm{{HF}}$, lifts the degeneracy of the nuclear energy levels and thus allows to perform NMR experiments without the need to apply an external static magnetic field. Under the action of the {\it rf}-field $h_{a}^{rf}$, the electronic magnetization oscillates and its angle of oscillation is given as $h_{a}^{rf}/\textrm{H}_{\textrm R}$, where $\textrm{H}_{\textrm{R}}$ is the restoring field acting upon the magnetic moments due to the various anisotropies present in the system~\cite{Panissod00,Panissod02}. The oscillation of the electronic magnetization is directly followed by the strong hyperfine field itself, which thus acquires an oscillating transverse component $h_{\textrm{HF}}^{rf}$. The angle of oscillation of the hyperfine field is given as $h_{\textrm{HF}}^{rf}/\textrm{H}_\textrm{{HF}}=h_{a}^{rf}/\textrm{H}_{\textrm{R}}$.\cite{Panissod00,Panissod02} The transverse component of the hyperfine field $h_{\textrm{HF}}^{rf}$ is larger than the applied {\it rf}-field $h_{a}^{rf}$ by the so-called enhancement factor $\eta$~\cite{Gossard59,Panissod00,Panissod02},
\be\label{eta}
\eta =\frac{h_{\textrm{HF}}^{rf}}{h_{a}^{rf}}=\frac{\textrm{H}_\textrm{{HF}}}{\textrm{H}_{\textrm{R}}}~.
\ee
Thus an independent measurement of the enhancement factor and of the hyperfine field gives direct access to the local restoring fields and allows one to obtain important information upon the magnetic stiffness in our samples.  In addition, this can be done for each separate line in the NMR spectrum and thus for each different magnetic or structural environment in the system.

\begin{figure}[t] 
\centering
\includegraphics[width=0.49\textwidth]{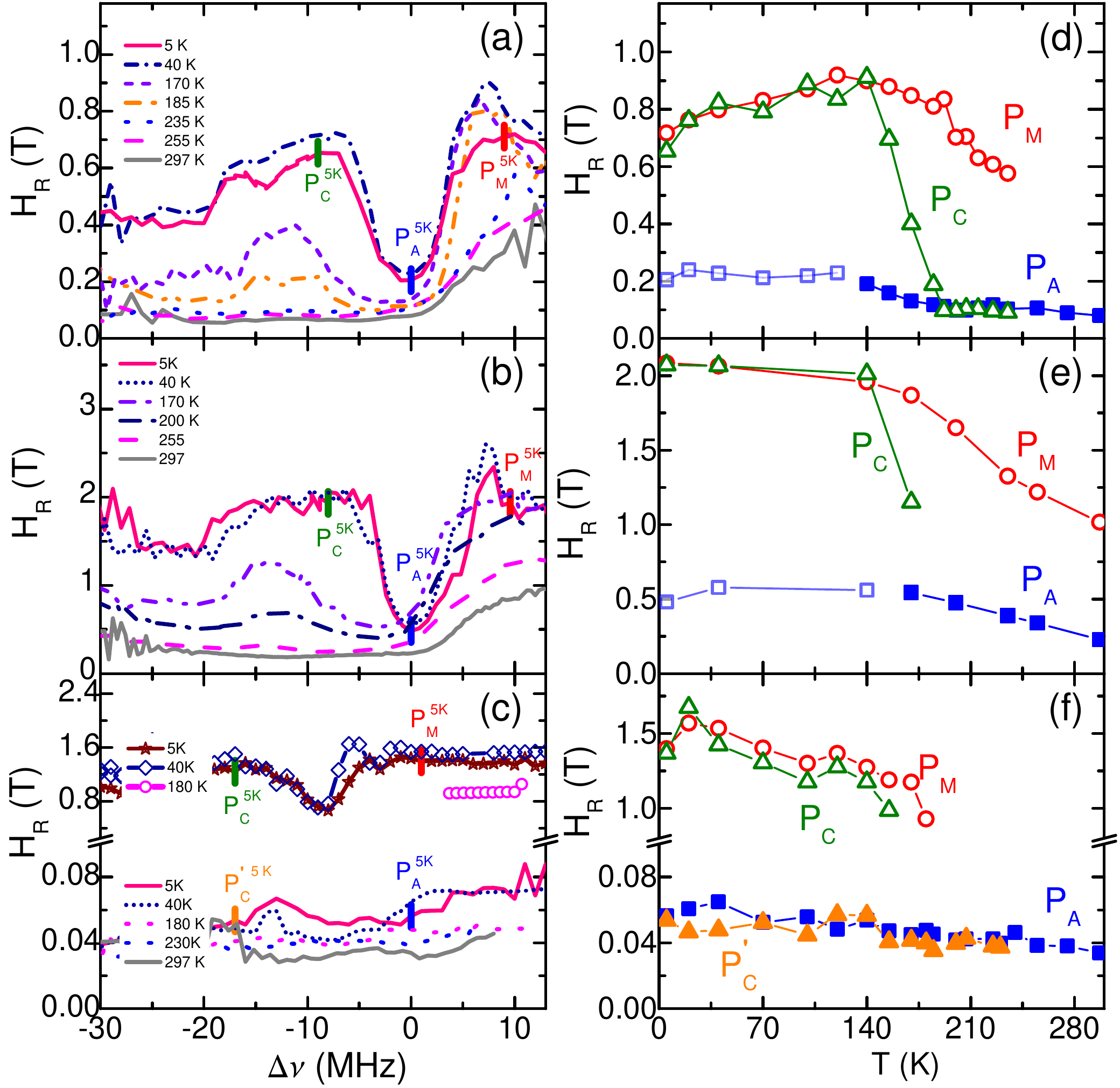} 
\caption{(Color online) The restoring field $\textrm{H}_{\textrm{R}}$ as a function of the frequency
shift $\Delta\nu=\nu-\nu_{o}$ at various temperatures for (a) sample-I, (b) sample-II and (c) sample-III. 
The reference point, $\nu_{o}$, is set to the position of the austenitic peak P$_A$ at high temperatures, while at low temperatures as the resonance frequency where the minimum restoring field is observed. In (d)-(f) we plot the temperature dependence of the restoring field for the austenitic peak P$_A$, the martensitic peak P$_M$ and the peaks P$_{\textrm{C}}$ and P$_{\textrm{C}}'$ for the three samples.}\label{AllRestoringFields2}
\end{figure}

The enhancement factor is also important upon signal reception. The precession of the nuclear magnetization drives the precession of the electronic magnetization, and so the NMR signal intensity I$(\omega)$ (the number of nuclei in each different frequency  $\omega$) is enhanced by the same factor $\eta$, as happens upon excitation.\cite{Gossard59,Panissod00,Panissod02} 
It has been shown that for a single phased ferromagnetic material, and due to the omnipresent distribution of enhancement factors, the observed NMR signal intensity S$(h_{a}^{rf},\omega)$ 
will tend to the log-normal distribution,\cite{Panissod00,Panissod02}
\be\label{S_omega}
\text S(h_{a}^{rf},\omega)= \eta(\omega) \text I(\omega)\text{exp}[-\text{log}^2(h_{a}^{rf}/h_{a,opt}^{rf})/2\sigma^2]~,
\ee
where $\sigma$ sets the width of the Gaussian distribution, and $h_{a,opt}^{rf}$ sets the value 
of the applied {\it rf}-field which gives the maximum signal intensity. This happens when the oscillating transverse 
component $h_{\textrm{HF}}^{rf}$ of the hyperfine field acquires the value $h_{\textrm{HF,opt}}^{rf}=\pi/2\gamma\tau$, 
where $\tau$ is the pulse length and $\gamma$ the gyromagnetic ratio.\cite{Panissod00,Panissod02} 

Due to the {\it rf}-enhancement mechanism the NMR sensitivity in ferromagnets is significantly improved. 
On the other hand special care should be taken in order to extract from the raw experimental data the 
actual number of nuclei in each different frequency I$(\omega)$. According to Eq.~(\ref{S_omega}), this requires to measure the signal intensity by varying the amplitude of $h_{a}^{rf}$ for each resonance frequency in the NMR spectrum.\footnote{We note here that the NMR spectra presented in Fig.~\ref{allLines} are corrected for the enhancement and thus the relative intensities are proportional to the number of resonating nuclei at each different frequency at time $2\tau$.}  These measurements give access to the value of $h_{\textrm{HF,opt}}^{rf}$ and, in turn, to the restoring field $\textrm{H}_{\textrm{R}}$ via Eq.~(\ref{eta}).

The frequency dependence of the restoring field $\textrm{H}_{\textrm{R}}$ is presented in Fig.~\ref{AllRestoringFields2}(a)-(c) 
for some representative temperatures, while in Fig.~\ref{AllRestoringFields2}(d)-(f) we show the T-dependence of the restoring fields for the main 
peaks observed in the NMR spectra. A first observation is that in samples I and II there is a frequency 
dependence of the restoring field, which implies the coexistence of magnetic environments with very distinct magnetic stiffness.
In this way, NMR gives access to valuable local information which is otherwise not accessible from e.g. bulk magnetization measurements, 
that provide only a weighted average of the anisotropy. 

The frequency dependence of the restoring field becomes more evident by lowering the temperature where, in addition, a significant gradual 
enhancement of the restoring field is observed. The enhancement of $\textrm{H}_{\textrm{R}}$ is signaling the transformation of an increasing number of 
austenitic regions to the low-T martensitic phase which has higher anisotropy. The dip observed in the 
frequency dependence of the restoring field for samples I and II down to 5 K (Fig.~\ref{AllRestoringFields2}(a), (b)) indicates that remnants of the high-T austenitic phase remain deep inside the martensitic phase. This behavior is supported by the fact that the resonance frequency of these regions is smoothly connected to the resonance frequency of the high-T austenitic phase as can be seen in Fig.~\ref{FvsTandFWHMall}, where the position of the dip is given by open symbols. We should also point out here that the amount of the low-T austenitic remnants is minute compared to the martensitic regions as can be seen in the intensity plots in Figs.~\ref{allLines}(a) and (b). 

Regarding the sample-II, we should emphasize that the values of restoring fields and thus the magnetic stiffness in this particular sample are higher compared to the other two samples. This occurs not only in the low-T martensitic phase but already from the high-T phase and indicates that the sample preparation process has created strains and precursors of the martensitic phase already at high-T. We also find that by annealing sample-II (Sec.~\ref{expdetails}), the strains and the martensitic remnants can be eliminated. As can be seen in Figs.~\ref{AllRestoringFields2}(c),(f), in sample-III the high-T phase is characterized by very low restoring fields and thus the annealing treatment has a large impact upon the magnetic stiffness. Furthermore, we note that the restoring field shows a weak frequency dependence at high-T in sample-III, showing a larger degree of magnetic homogeneity.

We also note that in sample-III and below 180 K, the NMR signal intensity $S(h_{a}^{rf},\omega)$ is described by a double Gaussian distribution in $h_{a}^{rf}$ instead of the high-T single Gaussian distribution of Eq.~(\ref{S_omega}). The high restoring field value for the new component (Fig.~\ref{AllRestoringFields2}(c)) shows that this component is martensitic.  The appearance of this martensitic component at 180 K in our NMR measurements is in agreement with the magnetic measurements in Fig.~\ref{SQUID}. Finally, remnants of the austenitic phase are observed down to 5 K but their intensity is negligible compared to the martensitic phase (Fig.~\ref{allLines}(c)).

\begin{figure}[!t] 
\centering
\includegraphics[width=0.47\textwidth]{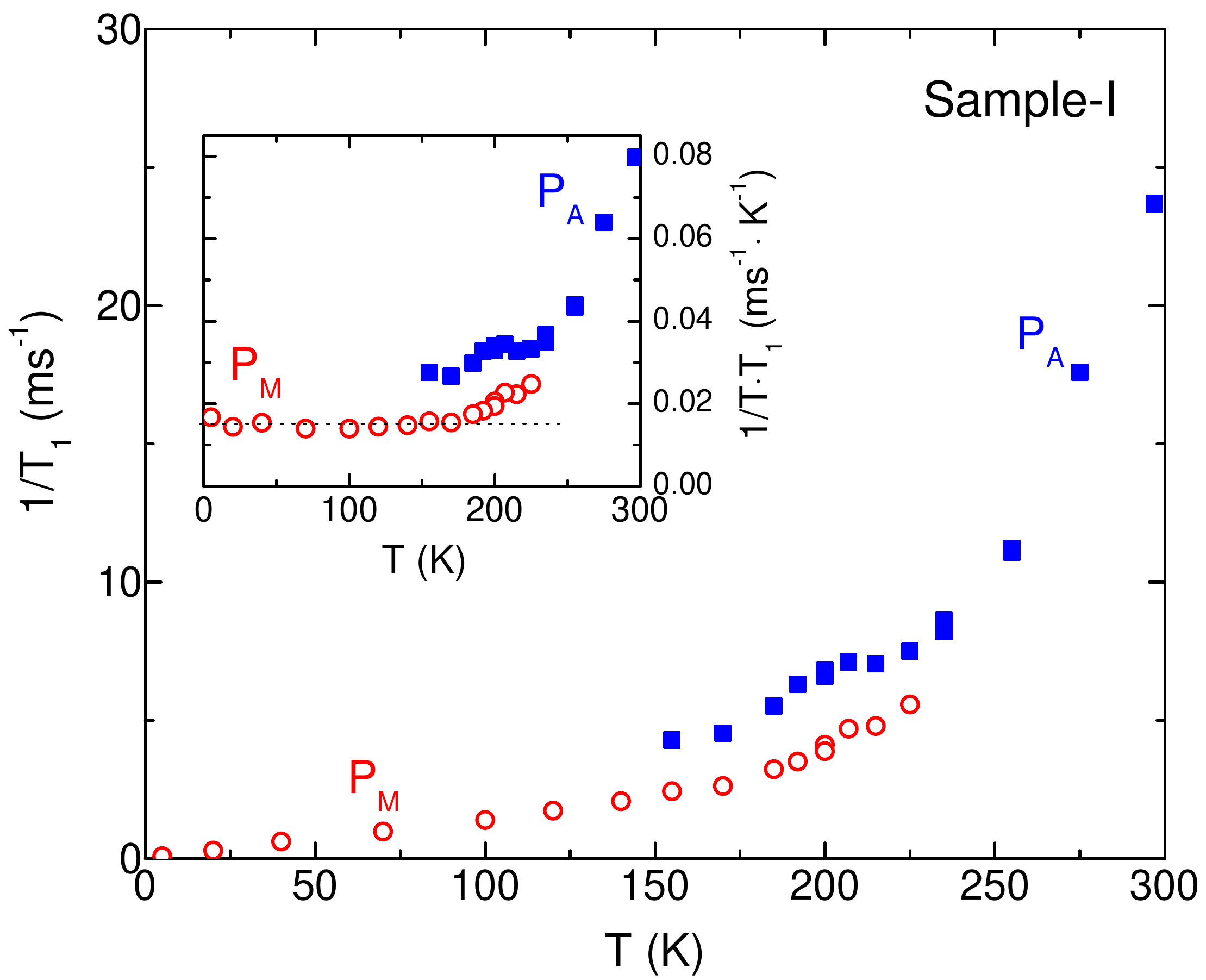} 
\caption{(Color online) Temperature dependence of $^{55}$Mn spin-lattice relaxation rate $1/T_1$ in the austenitic (squares)
and the martensitic (circles) phase of bulk (sample-I) Ni$_2$MnGa. The inset shows the $1/TT_1$ as a function of temperature.}
\label{T1bulkL}
\end{figure}

\subsubsection{Nuclear spin-lattice relaxation}\label{T1}
The spin-lattice relaxation rates $1/T_1$ in the austenitic and martensitic phases for sample-I are presented in Fig.~\ref{T1bulkL}.
At low temperatures, well inside the martensitic phase, the nuclear spin-lattice relaxation rate $1/T_1$ follows a linear-T dependence (see inset of Fig.~\ref{T1bulkL}). 
This behavior is expected for d-band FM metals where $1/T_1$ is dominated by fluctuating orbital and dipolar interactions due to electrons at the Fermi level and is given as $1/T_{1}\propto T\left[\rho_\uparrow\left(E_F\right)^2+\rho_\downarrow\left(E_F\right)^2\right]$, where $\rho_{\uparrow\left(\downarrow\right)}\left(E_F\right)$ the density of 
d-band states for up (down) spins at the Fermi level.\cite{Obata63,Moriya64,Kuhns06} A deviation from the linear temperature dependence is observed as we are approaching the martensitic transition, where the spin-lattice relaxation rate $1/T_1$ becomes faster.

In addition, we find that the spin-lattice relaxation rate $1/T_1$ is lower in the martensitic phase compared  
to the austenitic phase in the temperature region where the two phases coexist. This behavior indicates lower 
density of states at the Fermi level for the martensitic phase compared to the austenitic phase.
This result is supporting ultraviolet-photoemission (UPS) measurements which show redistribution in the intensity of the UPS spectra at both the premartenstic and martensitic transitions.\cite{Opeil08,Souza12}

\section{Summary}\label{sec:disc}
We have studied bulk and powder samples of the Heusler shape memory alloy Ni$_2$MnGa with zero-field static and dynamic $^{55}$Mn NMR experiments, X-ray powder diffraction and magnetization experiments. Besides the sequence of structural phase transitions in this compound, from the high-T austenitic phase down to the low-T martensitic phase, our NMR experiments also give access to the stiffness and local anisotropy for each separate magnetic environment via a detailed investigation of the so-called {\it rf}-enhancement factor. In doing this, we are able to differentiate signals coming from austenitic and martensitic components and follow their evolution with temperature. We also find that sample preparation has a strong impact on the weight of these components in each temperature region. Specifically, we show that powderization gives rise to a significant portion of martensitic traces inside the high-T austenitic region, and that these traces can be subsequently removed by annealing. Our X-ray measurements are in agreement with NMR and in addition show that the martensitic phase has orthorhombic structure with 7M periodicity.


\section{Acknowledgments}
We acknowledge experimental assistance from K. Leger, S. M\"{u}ller-Litvanyi, A. Omar and M. Gellesch and fruitful discussions with I. Rousochatzakis. 
S. W. acknowledges funding by Deutsche Forschungsgemeinschaft DFG in project WU 595/3-1.


\end{document}